\documentclass{aipproc}

\layoutstyle{6x9}

\usepackage{dsfont}
\usepackage{amsmath}
\usepackage{amssymb}
\usepackage{eucal}
\usepackage{mathrsfs}

\newcommand{\R}{\mathbb R}

\newcommand{\C}{\mathbb C}                           

\newcommand{\Z}{\mathbb Z}

\newcommand{\T}{\mathbb T}

\newcommand{\sss}[1]{\CMcal{#1}}
\newcommand{\bbb}[1]{\mathscr{#1}}

\newcommand{\num}[1]{\mathds {#1}}

\newcommand{\braket}[2]{\langle #1|#2\rangle}        

\newcommand{\expo}[1]{\mbox{{\upshape e}}^{#1}}                 

\newcommand{\ncint}{\mathrel{{\ooalign{$\int$\cr\kern+.07em\raise.15ex\hbox{$\pmb{\scriptstyle-}$}\cr}}}\!\!}

\newcommand{\ncpartial}{\mathrel{{\ooalign{$\partial$\cr\kern+.29em\raise.79ex\hbox{$\boldsymbol{\scriptstyle-}$}\cr}}}\!\!}
\newcommand{\ncC}{\mathrel{{\ooalign{$C_1$\cr\kern-.13em\raise.45ex\hbox{$\boldsymbol{\scriptstyle-}$}\cr}}}\!\!}

\newcommand{\virg}[1]{\lq\lq#1\rq\rq}                

\newcommand{\ii}{\,{\rm i}\,}
\def\dd{{\rm d}}

\begin{document}

\title{Topological aspects of generalized \\ Harper operators}


\author{Giuseppe De Nittis}{
address={D\'{e}partement de Math\'{e}matiques, Universit\'{e} de Cergy-Pontoise, 95302 Cergy-Pontoise, France},
email={denittis@math.univ-paris13.fr}
}

\author{Giovanni Landi}{
address={Dipartimento di Matematica, Universit\`{a} di Trieste, 
I-34127 Trieste, Italy\\ and INFN, Sezione di Trieste Italy},
email={landi@units.it}
}

\begin{abstract}
A generalized version of the TKNN-equations computing Hall conductances for generalized Dirac-like Harper operators is derived. Geometrically these equations relate Chern numbers of suitable (dual) bundles naturally associated to spectral projections of the operators.
\end{abstract}

\maketitle

\section{Generalized Harper operators}\label{se:intro}
The  \emph{integer quantum Hall effect} (IQHE) reveals a variety of 
surprising and attractive physical features, and has been the subject 
of several investigations 
(see \cite{morandi-88,graf-07} and references therein). 
In fact, a complete spectral analysis of the Schr\"{o}dinger operator for a single particle moving in a plane in a periodic potential and  subject to an uniform orthogonal magnetic field of strength $B$ (\emph{magnetic Bloch electron}) is extremely difficult. Thus the need for simpler {effective models} which hopefully capture (some of) the
main physical features in suitable physical regimes.

\smallskip

In the limit of a strong magnetic field, $B\gg1$, the IQHE is well described by an effective 
{Harper operator} (cf. \cite{wilkinson-87, bellissard-89, helfer-sjostrand-89, denittis-panati-10}). For this model the quantization (in units of  ${e^2}/{h}$) of the Hall conductance has a geometric meaning being related to Chern numbers of
suitable naturally bundles associated to spectral projections of the operator.  
A family of Diophantine equations, the {TKNN-equations} of \cite{Thouless-Kohmoto-Nightingale-Nijs-82}, provides a recipe for computing such integers. The aim of the present paper is to derive a generalized version of the TKNN-equations yielding the Hall conductances for more general {Dirac-like} Harper operators. 
Interest in such generalizations comes also from these
Dirac-like operators appearing naturally in important physical models, notably models for the graphene.

\smallskip

With $\theta= {1}/{B}$, the effective Harper operator is 
\begin{equation}\label{eq_001}
(H_{1,0}^\theta\psi)(x)= \psi(x-\theta)+\psi(x+\theta) + 2\cos(2\pi x)\psi(x) ,
\end{equation} 
acting on the Hilbert space $\sss{H}_1=L^2(\R)$. That this operator is the simplest representative of a large family of 
{generalized} Harper operators, sharing similar mathematical properties, is our starting point. 
On the Hilbert space $\sss{H}_1$ consider the unitary operators
\begin{equation}\label{eq_004}
(T_1 \psi)(x) =\expo{\ii 2\pi x}\psi\left(x\right) ,  \qquad 
(T^\theta_2 \psi)(x) =\psi\left(x-\theta\right),
\end{equation}
with $\theta\in\R$. They are readily seen to obey the relation 
\begin{equation}\label{nct}
T_1T_2^\theta=\expo{\ii 2\pi \theta}\ T_2^\theta T_1 ,
\end{equation} 
yielding for the Harper operator the expression $H_{1,0}^\theta=T_1+(T_1)^{\dag}+T_2^\theta+(T_2^\theta)^\dag$.

\smallskip

For any positive integer $q=1,2,\ldots,$ on the vector space $\C^q$ consider two unitary $q\times q$ matrices $\num{U}_q$ and $\num{V}_q$ defined as follows. Let $\{e_0,\ldots,e_{q-1}\}$ be the canonical basis of $\C^q$, 
then $\num{U}_q$ is a {diagonal matrix} and $\num{V}_q$ is a {shift matrix} acting as 
$$
\num{U}_q :e_{j}\mapsto\expo{\ii  2\pi\frac{j}{q}}\ e_{j} , \quad \textup{and} \quad 
\num{V}_q :e_{j}\mapsto e_{[j+1]_q} ,
$$ 
where $[\ \cdot\ ]_q$ stays for {modulo $q$}. They obey 
$$
\num{U}_q \num{V}_q = \expo{\ii  2\pi\frac{1}{q}} \num{V}_q \num{U}_q\qquad \rm{and}
 \qquad(\num{U}_q)^q= \num{I}_q=(\num{V}_q)^q.
$$ 

\smallskip

Then, on the Hilbert space  $\sss{H}_q=L^2(\R)\otimes\C^q$ one defines a pair of unitary operators:
\begin{equation}\label{eq_006}
U_q=T_1\otimes\num{U}_q, \qquad V^\theta_{q,r}=T_2^{\epsilon}\otimes(\num{V}_q)^r ,
\end{equation}
with $\epsilon(\theta,q,r)=\theta-\frac{r}{q}$ and $T_1$ and $T^\epsilon_2$ given by (\ref{eq_004}).  The integer $r\in\{0,\pm1,\ldots,\pm(q-1)\}$ is chosen {coprime} with respect to $q$. As for the  case before (when $q=1, r=0$), the operators (\ref{eq_006}) also obey the relation (\ref{nct}): 
$$
U_qV^\theta_{q,r}=\expo{\ii 2\pi \theta}\ V^\theta_{q,r}U_q .
$$
Following the definition (\ref{eq_001}) we can introduce the \emph{generalized $(q,r)$-Harper operator}
\begin{equation}\label{eq_001_bis}
H_{q,r}^\theta= U_q+(U_q)^{\dag}+V^\theta_{q,r}+({V^\theta_{q,r}})^{\dag} .
\end{equation}

\smallskip

More generally one considers the collection $\sss{A}_{q,r}^\theta$ of bounded operators on the Hilbert space $\sss{H}_q$  generated by the unitaries $U_q$  and $V^\theta_{q,r}$. Technically 
$\sss{A}_{q,r}^\theta$ is a $C^\ast$-algebra, i.e. an involutive algebra closed with respect to the operator norm, and it is named the \emph{(ir)rational rotation algebra} or the \emph{noncommutative torus algebra} \cite{rieffel-81,connes-rieffel-87}.  

\smallskip

By writing the operator in (\ref{eq_001}) as $H_{1,0}^\theta=D_\theta+C$ with 
\begin{equation}\label{eq_001bis}
(D_\theta\psi)(x)  =\psi(x-\theta)+\psi(x+\theta),\qquad
(C\psi)(x)  = 2\cos(2\pi x)\psi(x) ,
\end{equation}
in particular, the generalized $(2,1)$-Harper operator $H^\theta_{2,1}$ is just 
\begin{equation}\label{eq_003}
H^\theta_{2,1}=\left(
\begin{array}{cc}
C &  D_{\theta-\frac{1}{2}}\\ 
D_{\theta-\frac{1}{2}} & -C
\end{array}\right)
\end{equation}
acting on $\sss{H}_2=L^2(\R)\otimes\C^2$.  This operator provides an interesting effective model for the IQHE on {graphene}  \cite{bellissard-kreft-seiler-91,hatsugai-fukui-aoki-06,STK-08}.  
Moreover, {Dirac-like} operators like $H_{2,1}^\theta$  
can be used  to describe effective models for electrons  interacting with the periodic structure of a crystal
through a periodic (internal) magnetic field and subjected to the action of an external strong magnetic field  \cite{denittis-10, denittis-panati-10}.

\section{Bloch-Floquet transform}

For rational {deformation} parameter, $\theta={M}/{N}$ ($M$ and $N$ taken to be coprime here and after), any family of operators $\sss{A}_{q,r}^\theta$ can be decomposed in a continuous way according to a generalized version of the Bloch theorem. More explicitly we have the following.

\medskip

{\bf Proposition\ A.}\ \emph{
Let $\theta={M}/{N}$. For any (admissible) pair $(q,r)$ the bounded operator algebra $\sss{A}_{q,r}^\theta$ on $\sss{H}_q$
admits a \emph{bundle representation} $\Pi_{q,r}$ over the ordinary two-torus $\num{T}^2$. That is to say,  
there is a Hermitian vector bundle ${E}_{N,q}\to\num{T}^2$ together with a unitary transform 
$\bbb{F}_{q,r}:\sss{H}_q\to{L^2}({E}_{N,q})$ such that 
\begin{equation}\label{tildepi}
\Pi_{q,r}(\sss{A}_{q,r}^\theta)= 
\bbb{F}_{q,r}\ \sss{A}_{q,r}^\theta\ {\bbb{F}_{q,r}}^{-1}\subset\Gamma\big({\rm End}({E}_{N,q})\big) .
\end{equation}
The vector bundle ${E}_{N,q}$  has rank $N$ and (first) Chern number $C_1({E}_{N,q})=q$. 
}

\medskip

Here ${L^2}({E}_{N,q})$ denotes the Hilbert space of square integrable sections of the vector bundle ${E}_{N,q}$ and 
$\Gamma({\rm End}\big({E}_{N,q})\big)$ denotes the collection of {continuous} sections of the 
{endomorphism bundle} ${\rm End}({E}_{N,q})\to\num{T}^2$, i.e. the vector bundle with fibers 
${\rm End}(\C^q)$ associated with the vector bundle ${E}_{N,q}$.
The unitary map $\bbb{F}_{q,r}$ implementing the bundle representation of $\sss{A}_{q,r}^\theta$ is called 
\emph{(generalized) Bloch-Floquet transform} \cite{denittis-10, denittis-panati-09}. 
For the details of the  proof of Prop.~A, 
that we briefly sketch, 
we refer to \cite{denittis-landi-11} (see also \cite{rieffel-83}).
Denote with $\alpha\in\Z$, $|\alpha|<q$ the unique solution 
of $\beta q-\alpha r=1$ (due to $q$ and $r$ being coprime) and be $M_0=qM-rN$. Then, a simple check shows that the unitary operators 
$$ 
A_{q,r}^\theta=(T_1)^{\frac{1}{q\epsilon}}\otimes(\num{U}_q)^{\alpha}, \qquad B_{q,r}^\theta=T_2^{\frac{M_0}{q}}\otimes(\num{V}_q)^{rN}
$$
commute, $[A_{q,r}^\theta, B_{q,r}^\theta]=0$, while commuting with any element in $\sss{A}_{q,r}^\theta$. 
They generate a (indeed maximally) commutative sub-algebra of the commutant of $\sss{A}_{q,r}^\theta$, and in particular, of symmetries for the operator (\ref{eq_001_bis}). 
Were $N$ a multiple of $q$ this commutative sub-algebra would reduce to a direct sum of $q$ copies of a commutative algebra on $L^2(\R)$. 
Thus to avoid this degeneracy, we take $q$ and $N$ to be coprime as well. 
This entails there exist two integers $d_r$ and $n_r$ such that 
$q d_r + N n_r = 1$, a fact we shall exploit momentarily.  Moreover, the commutative algebra generated by 
$A_{q,r}^\theta$ and $B_{q,r}^\theta$ is isomorphic to the algebra of continuos functions over the ordinary 
2-torus $\num{T}^2$.

\smallskip

A generalized simultaneous eigenvectors of $A_{q,r}^\theta$ and $B_{q,r}^\theta$ is a $\Xi_k\in S'(\R)\otimes\C^q$ ($S'(\R)$ is the space of tempered distributions) such that
$$
A_{q,r}^\theta\ \Xi_k=\expo{\ii 2\pi k_1}\ \Xi_k,\qquad B_{q,r}^\theta\ \Xi_k=\expo{\ii 2\pi k_2}\ \Xi_k.
$$
For any $k=(k_1,k_2)\in[0,1]^2\simeq\T^2$, the generalized eigenvectors make up a $N$-dimensional space, a basis of which being given by a fundamental family of distribution $\Upsilon^{(j)}(k)=(\zeta_0^{(j)}(k),\ldots,\zeta_{q-1}^{(j)}(k))\in S'(\R)\otimes\C^q$, for indices $j=0,\ldots,N-1$, with elements $\zeta^{(j)}_\ell(k)$, $\ell=0,\ldots,q-1$, defined by
 \begin{equation}\label{eq_sec_har_01bis}
\zeta^{(j)}_\ell(k)=\sqrt{\frac{|M_0|}{N}}\ \sum_{m\in\Z}\expo{- \ii2\pi k_1(\tau_\ell+mq)}
\ \delta\left[\ \cdot\ -\frac{M_0}{N}(k_2+j) -mM_0-\tau_\ell\frac{M_0}{q}\right] .
\end{equation} 

\smallskip

Here the permutation $\tau:\ell\mapsto\tau_\ell$ of the set $\{0,\ldots,q-1\}$ is defined  by $\ell=[\tau_\ell rN]_q$
and, as usual, the Dirac delta function $\delta(\cdot-x_0)$ acts on functions $f:\R\to\C$ as the evaluation at the point $x_0$, i.e. $\langle \delta(\cdot-x_0);f\rangle=f(x_0)$.

\smallskip

We let $\sss{H}_{q,r}(k)\subset S'(\R)\otimes\C^q$ denote the $N$-dimensional vector space spanned by the distributions $\Upsilon^{(0)}(k),\ldots,\Upsilon^{(N-1)}(k)$. The total space of the vector bundle ${E}_{N,q}$ is just the disjoint union of the spaces $\sss{H}_{q,r}(k)$ glued together with transitionf functions coming from pseudo-periodic conditions satisfied by the $\Upsilon^{(j)}$'s. Indeed, from (\ref{eq_sec_har_01bis}) one deduces that 
$\Upsilon^{(j)}(k_1+1,k_2)=\Upsilon^{(j)}(k_1,k_2)$ while  $\Upsilon^{(j)}(k_1,k_2+1)=\Upsilon^{(j+1)}(k_1,k_2)$ for $j=0,\ldots,N-2$ and  $\Upsilon^{(N-1)}(k_1,k_2+1)=\expo{\ii 2\pi q}\Upsilon^{(0)}(k_1,k_2)$. 
Also, there is an identification
$$
\bbb{F}_{q,r}:\sss{H}_q\to{L^2}({E}_{N,q})\simeq\int_{\T^2}^\oplus\sss{H}_{q,r}(k)\ \dd z(k)
$$
which is very reminiscent of the usual direct integral decomposition of the Bloch theory. We stress that  
Prop.~A 
does not only states that  any $H\in\sss{A}_{q,r}^\theta$ can be decomposed as a direct integral operator $H=\int_{\T^2}^\oplus h(k)\ \dd z(k)$ with $h(k)$ an $N\times N$ matrix acting on $\sss{H}_{q,r}(k)$,  but also that such a decomposition is continuous with respect to the topology of the vector bundle ${E}_{N,q}$, thus amounting to a bundle representation. For $H\in \sss{A}_{q,r}^\theta$, we denote $\widetilde{H}=\Pi_{q,r}(H)$. For the generators, when acting on the basis $\{ \Upsilon^{(j)}(k) \}$ one finds
\begin{equation}\label{matu-v}
\widetilde{U}_q(k_1,k_2) = \expo{\ii 2\pi\frac{M_0}{N}k_2 }\ (\num{U}_N)^{qM},\qquad
\widetilde{V}_{q,r}^{\theta}(k_1,k_2)  = \expo{\ii 2\pi n_r k_1}\ (\num{V}_{N;k_1})^{d_r}.
\end{equation}
Here $\num{U}_N$ is the diagonal matrix $\num{U}_N :e_{j}\mapsto\expo{\ii  2\pi\frac{j}{N}}\ e_{j}$;
$\num{V}_{N;k_1}$ is the {twisted} shift matrix sending 
$e_{j}$ to $e_{j+1}$ for $j=0, \dots, N-2$ while  $e_{N-1}$ to $\expo{\ii 2\pi q k_1} e_{0}$. 
The matrices in (\ref{matu-v}) commute up to $\expo{\ii 2\pi\frac{M}{N}qd_r}=\expo{\ii 2\pi\frac{M}{N}}$ being 
$qd_r=1-n_rN$ as before, 
$\widetilde{U}_q \widetilde{V}_{q,r}^{\theta} = \expo{\ii 2\pi\frac{M}{N}} \widetilde{V}_{q,r}^{\theta} \widetilde{U}_q$,
thus providing a representation of $\sss{A}_{q,r}^\theta$. Moreover, their pseudo-periodic conditions
$$
\widetilde{U}_q(k_1+1,k_2+1)=\expo{\ii2\pi\frac{M}{N}} \widetilde{U}_q(k_1,k_2), \qquad
\widetilde{V}_{q,r}^{\theta}(k_1+1,k_2+1)=\widetilde{V}_{q,r}^{\theta}(k_1,k_2) , 
$$
match those of the basis $\{ \Upsilon^{(j)}(k) \}$ thus making 
$\Pi_{q,r}$ a representation of  $\sss{A}_{q,r}^\theta$ as bundle endomorphisms, as expressed in \eqref{tildepi}.

\smallskip

The bundle ${E}_{N,q}$ comes equipped with the  \emph{Berry connection} 
\begin{equation}\label{eq_berry_con}
\omega_{i,j}(k)=
 \braket{\Upsilon^i(k)}{\, \dd \Upsilon^j(k)}\ , \quad i,j=0,\ldots,N-1 ,
\end{equation}
Its curvature $K= \dd \omega$ is  constant,
$K(k) =\left(\frac{ 2\pi q}{\ii N}\ \num{I}_N\right) \dd k_1\wedge \dd k_2$ (up to an exact form) and
when integrated it results in the first Chern number of the bundle being 
$$
C_1(E_{N,q})=\frac{\ii}{2\pi} \int_{\num{T}^2}{\rm Tr}_N( K)  = q .
$$

\section{Generalized TKNN-equations}

For a rational $\theta={M}/{N}$,
the spectrum of $H_{q,r}^\theta$ in (\ref{eq_001_bis})
has $N+1$ energy bands if $N$ is odd or $N$ energy bands if $N$ is even \cite{hofstadter-76, Choi-elliott-yui-90, denittis-landi-11}. These include the \emph{inf-gap} (from $-\infty$ to the minimum of the spectrum) and the \emph{sup-gap} (from the maximum of the spectrum to $+\infty$). 

\smallskip

To each gap $g$ one associates a spectral projection $P_g$
with the convention that $P_0=0$ for the inf-gap $g=0$ and $P_{\rm{max}}=\num{I}$ for the sup-gap $g=N_{\rm{max}}$ with $N_{\rm{max}}=N-1$ or $N_{\rm{max}}=N$ according to whether $N$ is odd or even. 
As usual, the projection $P_g$ is  defined via the Riesz formula for the operator $H_{q,r}^\theta$,  
\begin{equation}\label{rf}
P_g=\frac{1}{\ii 2\pi}\oint_\Lambda(\lambda\num{I}-H_{q,r}^\theta)^{-1}\ \dd \lambda .
\end{equation}
The closed rectifiable path  $\Lambda\subset\C$  
encloses the spectral subset  $I_g=[\varepsilon_0,\varepsilon_g]\cap\sigma(H_{q,r}^\theta)$ (intersecting the real axis in  $\varepsilon_0$ and $\varepsilon_g$) with the real numbers 
$\varepsilon_0,\varepsilon_g\in\R\setminus \sigma(H_{q,r}^\theta)$ being such that 
$-\infty<\varepsilon_0<\rm{min}\ \sigma(H_{q,r}^\theta)$ and $\varepsilon_g$ in the gap $g$. 

\smallskip

The Hall conductance associated with the energy spectrum up to the gap $g$ 
is related to the projection $P_g$ via the {Kubo formula} (linear response theory). 
Its value is an integer number $t_g$; it is by now well known that $t_g$ is to be 
thought of as the Chern number of a bundle determined by the projection $P_g$ \cite{Thouless-Kohmoto-Nightingale-Nijs-82, Bellissard-Baldes-Elst-94, avron-04}.  

\smallskip

Any such a spectral projection $P_g$ yields a projection $\Pi_{q,r}(P_g)\in 
\Gamma({\rm End}({E}_{N,q}))$, via the representation $\Pi_{q,r}$ in (\ref{tildepi}), and thus a vector subbundle $L_{q,r}(P_g)\subset {E}_{N,q}$. The related (first) Chern number $C_1(L_{q,r}(P_g))$ measures the degree of non triviality of the bundle $L_{q,r}(P_g)$. The geometric interpretation of the Hall conductance is none other than the equality 
$t_g=C_1(L_{q,r}(P_g))$. On the other hand, the Chern number $C_1(L_{q,r}(P_g))$ obeys a Diophantine equation which then provides a TKNN-type equation for the conductance $t_g$. We have the following.

\medskip

{\bf Proposition\ B.}\ \emph{
For any projection $P$ in the algebra $\sss{A}_{q,r}^{\theta}$ there exists a 
\virg{dual} vector bundle 
${L}_{\rm{ref}}(P)\to\num{T}^2$ s.t.
the following duality between Chern numbers holds:
\begin{equation}\label{geom_TKNN}
C_1(L_{q,r}(P)) = q\left[ \frac{1}{N}\ {\rm Rk}({L}_{{\rm ref}}(P)) + \left(\frac{M}{N}-\frac{r}{q}\right) C_1({L}_{{\rm ref}}(P)) \right].
\end{equation}
}

\smallskip

Before we sketch the proof of this result we turn to its interpretation in terms of conductances of the generalized Harper operators in (\ref{eq_001_bis}). As mentioned, if $P_g$ is its spectral projection up to the gap $g$, the associated Hall conductance $t_g$ is the number $C_1(L_{q,r}(P_g))$. For the dual number we have 
$C_1({L}_{\rm{ ref}}(P_g))=-s_g$, with $s_g$ identified with the Hall conductance of the energy spectrum up to the gap $g$ but in the opposite limit of a weak magnetic field ($B\ll1$) \cite{Thouless-Kohmoto-Nightingale-Nijs-82, avron-04, 
denittis-10}. Writing $d_g={\rm Rk}({L}_{{\rm ref}}(P_g))$, 
relation (\ref{geom_TKNN}) translates to the
\emph{generalized} TKNN-\emph{equations} 
\begin{equation}\label{gen-TKNN-eqs}
N\ t_g+(qM-rN)\ s_g = q \ d_g, \quad\quad g=0,\ldots,N_{\rm{max}}.
\end{equation}
When $q=1$ and $r=0$, the above reduces to
\begin{equation}\label{eq_002}
N\ t_g+M\ s_g=d_g , \quad\qquad g=0,\ldots,N_{\rm{max}}, 
\end{equation}
which is the original TKNN-equation derived in \cite{Thouless-Kohmoto-Nightingale-Nijs-82} for the Harper operator (\ref{eq_001}). In its spirit then,
the integer $d_g$ in the right-hand side coincides with the labeling of the gap when $N$ is odd, i.e.  $d_g=g$ for $N$ odd. When $N$ is even $d_g=g$ if $0\leqslant g\leqslant {N}/{2}-1$ and $d_g=g+1$ if ${N}/{2}\leqslant g\leqslant N_{\rm{max}}=N-1$. We remark that the bound 
\begin{equation}\label{eq_010bis}
2\, | s_g  |<N
\end{equation}
(already present in \cite{Thouless-Kohmoto-Nightingale-Nijs-82}) still holds, owing to the bound $2\, |C_1({L}_{\rm{ ref}}(P_g))|<N$ for the spectral projections into the gaps of the Hofstadter operator \cite{Choi-elliott-yui-90}.

\smallskip

Now, the subbundle $L_{q,r}(P)\subset E_{N,q}$ determined by the projection valued section 
${\Pi}_{q,r}(P)=P(\cdot)$, for  a projection $P\in\sss{A}_{q,r}^{\theta}$, will have as fiber over $k\in\num{T}^2$ the space 
\begin{equation}\label{fib-can=qr}
L_{q,r}(P) |_{k} = {\rm Range}(P(k))\subset \sss{H}_{q,r}(k) .
\end{equation}
We need a dual bundle representation, $\Pi_{q,r}^{\rm{ref}}$ of $\sss{A}_{q,r}^{\theta}$, s.t.
\begin{equation}\label{eq_proofGD4}
P(k_1,Nk_2)= P^{\rm{ref}}(k_1,M_0k_2), 
\end{equation}
and $P^{\rm{ref}}(\cdot)=\Pi_{q,r}^{\rm{ref}}(P)$.
We are lead to the representation
\begin{equation}\label{eq_ref_rep}
U^{\rm{ref}}_q(k) =\expo{\ii 2\pi k_2} \, (\num{U}_N)^{qM},\qquad
V^{\rm{ref}}_{q,r}(k) =V_{q,r}^{\theta}(k)=\expo{\ii2\pi n_r k_1} \left( \num{V}_{N;k_1}          
\right)^{{d_r}}. 
\end{equation}
which obey $U^{\rm{ref}}_q(\cdot) V^{\rm{ref}}_{q,r}(\cdot) = \expo{\ii2\pi\frac{M}{N}} V^{\rm{ref}}_{q,r}(\cdot) U^{\rm{ref}}_q(\cdot)$. As elements in $C(\num{T}^2)\otimes\rm{Mat}_N(\C)\simeq C(\num{T}^2;\rm{Mat}_N(\C))$ they yield a representation of  $\sss{A}_{q,r}^{\theta}$  as endomorphisms of the the trivial bundle $\num{T}^2 \times \C^N\to\num{T}^2$. 
Then, any projection $P$ in $\sss{A}_{q,r}^{\theta}$ is mapped to a projection-valued section
$P^{\rm{ref}}(\cdot)=\Pi_{q,r}^{\rm{ref}}(p)$ which defines a vector subbundle ${L}_{\rm{ref}}(p)\to\num{T}^2$ of the trivial vector bundle $\num{T}^2 \times \C^N$. It will   have as fiber over $k\in\num{T}^2$ the space 
\begin{equation}\label{fib-ref}
{L}_{\rm{ref}}(P) |_{k}  = {\rm Range}(P^{\rm{ref}}(k))\subset \C^N .
\end{equation}
Then, equation (\ref{eq_proofGD4}) say that the vector bundle $L_{q,r}({P})$ \virg{winded} around $N$ times in the second direction is (locally) isomorphic to the vector bundle $L_{\rm{ref}}({P})$ \virg{winded} around $M_0$ times in the same direction. There is however an extra twist, due to the bundle $E_{N,q}$, of which $L_{q,r}(P)$ is a subbundle, being not trivial.  Indeed, an analysis of the transition functions lead to the bundle isomorphism
\begin{equation}\label{rvb}
 \varphi_{(1,N)}^\ast L_{q,r}(P) \simeq \varphi_{(1,M_0)}^\ast{L}_{\rm{ref}}(P)\otimes{\rm det}({E}_{N,q}).
\end{equation}
Here ${\rm det}({E}_{N,q})\to \num{T}^2$ is the {determinant line bundle} and the extra operation $\varphi_{(1,N)}^\ast$ (the {pullback}) stays for the extra winding by $N$ (for the bundle $L_{q,r}({P})$) and the same for $\varphi_{(1,M_0)}^\ast$ 
(for the bundle $L_{\rm{ref}}({P})$). 
Formula (\ref{geom_TKNN}) is the relation among corresponding first Chern numbers. 
Using the fact that
$C_1(\varphi_{(1,N)}^\ast L_{q,r}(P)) = N C_1(L_{q,r}(P))$ and 
$C_1(\varphi_{(1,M_0)}^\ast{L}_{\rm{ref}}(P))=M_0 C_1({L}_{\rm{ref}}(P))$, 
as well as the identity $C_1({\rm det}({E}_{N,q}))=C_1({E}_{N,q})=q$, the relation (\ref{geom_TKNN}) follows from (\ref{rvb}) by standard arguments.

\section{The irrational case}
On the algebra $\sss{A}_{q,r}^\theta$ there is a faithful trace defined by 
$$
\tau \left( (U_q) ^n  (V^\theta_{q,r})^m \right)=\delta_{n,0}\ \delta_{m,0}
$$ 
on monomials, and extended by linearity. Derivations $\partial_j: \sss{A}_{q,r}^\theta\to\sss{A}_{q,r}^\theta$, for $j=1,2$,  defined on monomials by  
$$
\partial_1((U_q)^n (V^\theta_{q,r})^m)=\ii 2\pi\ n\ (U_q)^n (V^\theta_{q,r})^m, \qquad
\partial_2((U_q)^n (V^\theta_{q,r})^m)=\ii 2\pi\ m\ (U_q)^n (V^\theta_{q,r})^m ,
$$
are extended by linearity and Leibniz rule. 
Lastly, we need the first Connes-Chern number which, for a projection 
$P\in\sss{A}_{q,r}^\theta$ (in the domain of the derivations) computes the integer (an index of a Fredholm operator)
$$
\ncC(P)
=\frac{1}{\ii 2\pi}\tau \big( P (\partial_1(P)
\partial_2(P)-\partial_2(P)\partial_1(P) )  \big).
$$
Let $H_{q,r}^\theta \in\sss{A}_{q,r}^\theta$ be the Hofstadter operator 
(\ref{eq_001_bis}) with associated spectral projection $P_g^\theta$ for the gap $g$ as in (\ref{rf}). 
For $\theta\in I \subset \R$, the functional expression of $H_{q,r}^\theta \in\sss{A}_{q,r}^\theta$ 
is fixed and $H_{q,r}^\theta$ 
depends on the parameter $\theta$ only through the fundamental commutation relation  which defines $\sss{A}_{q,r}^\theta$.  Now, 
if the gap $g$ is open for all $\theta\in I$ (with $I$ sufficiently small),
the functions $\theta\mapsto\ \ncC(P_g^\theta)$
is constant in the interval $I$  \cite{boca-01}.
On the other hand, from the structure of the group $K_0(\sss{A}_{q,r}^\theta)$, one deduces \cite{pimsner-voiculescu-80, connes-94} that 
\begin{equation}\label{eq_c10b-bis}
\tau(P_g^\theta)=m\ P_g^\theta - \theta \ncC(P_g^\theta) ,
\end{equation}
with the integer $m(\cdot)\in\Z$ uniquely determined by the condition
$0\leqslant\tau(\cdot)\leqslant1$. From (\ref{eq_c10b-bis}), the integer $m(\cdot)$ is constant for $\theta\in I$. Hence,
 formula
\begin{equation}\label{eq_11}
C_{q,r}(P_g)  = q\left[m(P_g)-\frac{r}{q} \ncC(P_g)\right]  = q\left[\tau(P_g)+\left(\theta-\frac{r}{q}\right) \ncC(P_g)\right]\in\Z
\end{equation}
is well defined  and extends (\ref{geom_TKNN}) for irrational values $\theta\in I$ 
(for which the gap $g$ remains open). 
Indeed, for a rational $\theta={M}/{N}$ one has natural identifications 
$\rm{Rk}({L}_{\rm{ ref}}(P))=\tau(P)$ and $C_1({L}_{\rm{ref}}(P))=\ \ncC(p)$
\cite{denittis-landi-11} and for the rational torus, formula (\ref{eq_11}) is the same as (\ref{geom_TKNN}). 

\smallskip

We think of (\ref{eq_11}) as relating conductances for the Harper operator $H_{q,r}^\theta$ 
in (\ref{eq_001_bis}), thus generalizing (\ref{gen-TKNN-eqs}) to 
\begin{equation}\label{irr-TKNN-eqs}
 t_g+(q \theta - r) s_g = q \, d_g, 
\end{equation}
with $P_g$ once again the spectral projections of the Harper operator $H_{q,r}^\theta$, and now identifying 
$t_g=C_{q,r}(P_g)$  and $s_g= -\ncC(P_g)$ as before, whereas $d_g =\ \tau(P_g)$.

\begin{theacknowledgments}
 G.D. is supported by the grant ANR-08-BLAN-0261-01. G.L. was partially supported by the Italian Project \virg{Cofin08 -- Noncommutative Geometry, Quantum Groups and Applications}.  GL thanks Noureddine Mebarki and Jamal Mimouni 
for the kind invitation to the \virg{8th International Conference on Progress in Theoretical Physics} held at Mentouri University, Constantine, Algeria, October 2011. He is grateful to them and to all participants for the great time in Constantine.
\end{theacknowledgments}

\bibliographystyle{aipproc}   

\end{document}